\documentclass{appolb}
\usepackage{graphicx}
\usepackage{epsfig}
\usepackage{wrapfig}
\usepackage{subfig}

\begin{document}
\eqsec  

\title{ J/$\psi$ polarization in $p+p$ collisions \\at $\sqrt{s}$ = 200 GeV in STAR %
\thanks{Presented at Strangeness in Quark Matter,  18-24 September 2011 Polish Academy of Arts and Sciences, Cracow, Poland}%
}
\author{Barbara Trzeciak for the STAR Collaboration
\address{Faculty of Physics, Warsaw University of Technology \\ Koszykowa 75, 00-662 Warsaw, Poland}
}
\maketitle

\begin{abstract}
In this paper, J/$\psi$ polarization at mid-rapidity in $p+p$ collisions at \\$\sqrt{s}$ = 200 GeV measured in the STAR experiment at RHIC is reported. J/$\psi$ production is analyzed via the dielectron decay channel. J/$\psi$ polarization is extracted from the decay angular distribution in the helicity frame. The J/$\psi$ polarization is measured at transverse momentum range (2 - 6) GeV/$c$ and is found to be consistent with $NLO^+$ Color Singlet Model ($NLO^+$ CSM), Color Octet Model (COM) predictions and with no polarization within current uncertainties.
\end{abstract}
  
\section{Introduction}
A number of models with different J/$\psi$ production mechanisms are able to describe the measured J/$\psi$ production cross section reasonably well. It suggests that other observables are needed to discriminate between different J/$\psi$ production models.  J/$\psi$ spin alignment, commonly named as J/$\psi$ polarization, can be used as such an observable since various models have different $p_{T}$ dependent predictions on the J/$\psi$ polarization. 

The Color Evaporation Model (CEM) describes J/$\psi$ cross sections measured in many experiments reasonably well but has no prediction power regarding  the J/$\psi$ polarization. The $NLO^+$ Color Singlet Model (CSM) \cite{Lansberg} predicts longitudinal J/$\psi$ polarization in the helicity frame at low and mid $p_{T}$ at mid-rapidity.  That recent calculations of the CSM for the yield differential in $p_{T}$ show better agreement  with RHIC data at low and mid $p_{T}$ than earlier CSM calculations and the result for  J/$\psi$ polarization is in good agreement with the PHENIX data. Non-relativistic quantum chromodynamics (NRQCD) effective theory, also known as Color Octet Model (COM) predicts transverse polarization at high  J/$\psi$ $p_{T}$, $p_{T}>$ 5 GeV/$c$ which is in disagreement with the polarization measurement from the CDF experiment at FermiLab \cite{CDF}. On the other hand a tuned NRQCD predicts longitudinal  J/$\psi$ polarization at 1.5 $< p_{T}  <$ 5 GeV/$c$ at mid-rapidity and is able to qualitatively describe the PHENIX J/$\psi$  polarization measurements as well as the cross section measurements \cite{NRQCD}.

The measurement of J/$\psi$ polarization at $p_T > $ 5 GeV/$c$ is expected to have discrimination power against different models of  J/$\psi$ production since, e.g., $NLO^+$ CSM and COM predict different polarization in that  J/$\psi$ $p_T$ region.

\subsection{Decay angular distribution}
In this study,  J/$\psi$ polarization is analyzed via the angular distribution of the electron decay from charmonium in the helicity frame \cite{HX}. The angular distribution is derived from the density matrix elements of the production amplitude using parity conservation rules. Polar angle $\theta$ is the angle between the positron momentum vector in the  J/$\psi$ rest frame and  J/$\psi$ momentum vector in the laboratory frame.

The angular  distribution integrated over the azimuthal angle is parametrized: 
\begin{equation}
\frac{dN}{dcos\theta} \propto 1+\lambda cos^2\theta
\label{parametrization}
\end{equation}
where $\lambda$ is the polarization parameter that contains both the longitudinal and transverse components of the  J/$\psi$ cross section. When $\lambda$ = 0 there is no polarization, $\lambda$ = -1 means full longitudinal polarization and $\lambda$ = 1 corresponds to full transverse polarization.

\section{Data analysis}
In this analysis, data recorded in 200 GeV $p+p$ collisions in the STAR experiment in year 2009 is used. The analyzed data was sampled from an integrated luminosity of $\sim$1.5 $pb^{-1}$ and was triggered by the STAR Barrel Electromagnetic Calorimeter (BEMC). The trigger required transverse energy deposited in a single BEMC tower ($\Delta \eta \times \Delta \phi = 0.05 \times 0.05$) to be within 2.6 $< E_{T}  \leq$ 4.3 GeV. 

J/$\psi$ is reconstructed via its dielectron decay channel  J/$\psi \rightarrow e^+ e^-$  with branching ratio BR = 5.9\%, and is required that at least one of electrons from J/$\psi$ decay satisfies the trigger conditions. The Time Projection Chamber (TPC), Time Of Flight (TOF) and BEMC detectors are used to reconstruct and identify electrons.
The TPC provides information about dE/dx. Information from the TOF is very useful for electron identification and hadron rejection at lower momenta, where electron and hadron dE/dx bands overlap, $\mid1/\beta -1\mid \; < \;$ 0.03 ($\beta$ = v/$c$) cut was applied at p $<$ 1.4 GeV/$c$.  In 2009, 72\% of the TOF detector was installed. At higher momenta the BEMC can reject hadrons very efficiently. For momenta above 1.4 GeV/$c$ a cut of E/p $>$ 0.5 was used, where E is energy deposited in a single BEMC tower (for electrons E/p ratio is expected to be $\approx$ 1).

\subsection{J/$\psi$ signal}
\begin{figure}[!ht]
 \vspace{-20pt}
\centering
\subfloat[]{\label{mass} \includegraphics[scale=0.25]{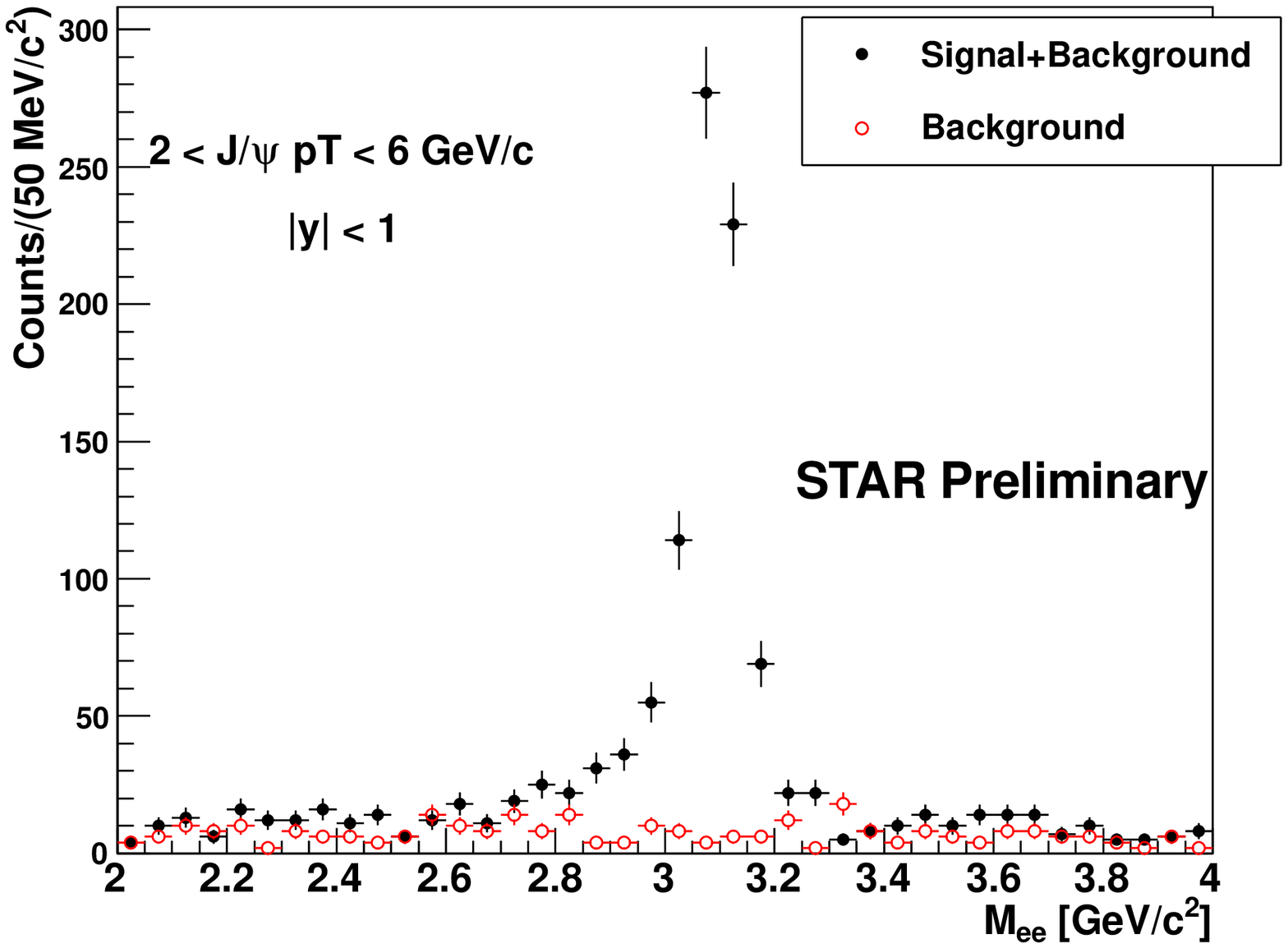} }
\subfloat[]{\label{signal} \includegraphics[scale=0.25]{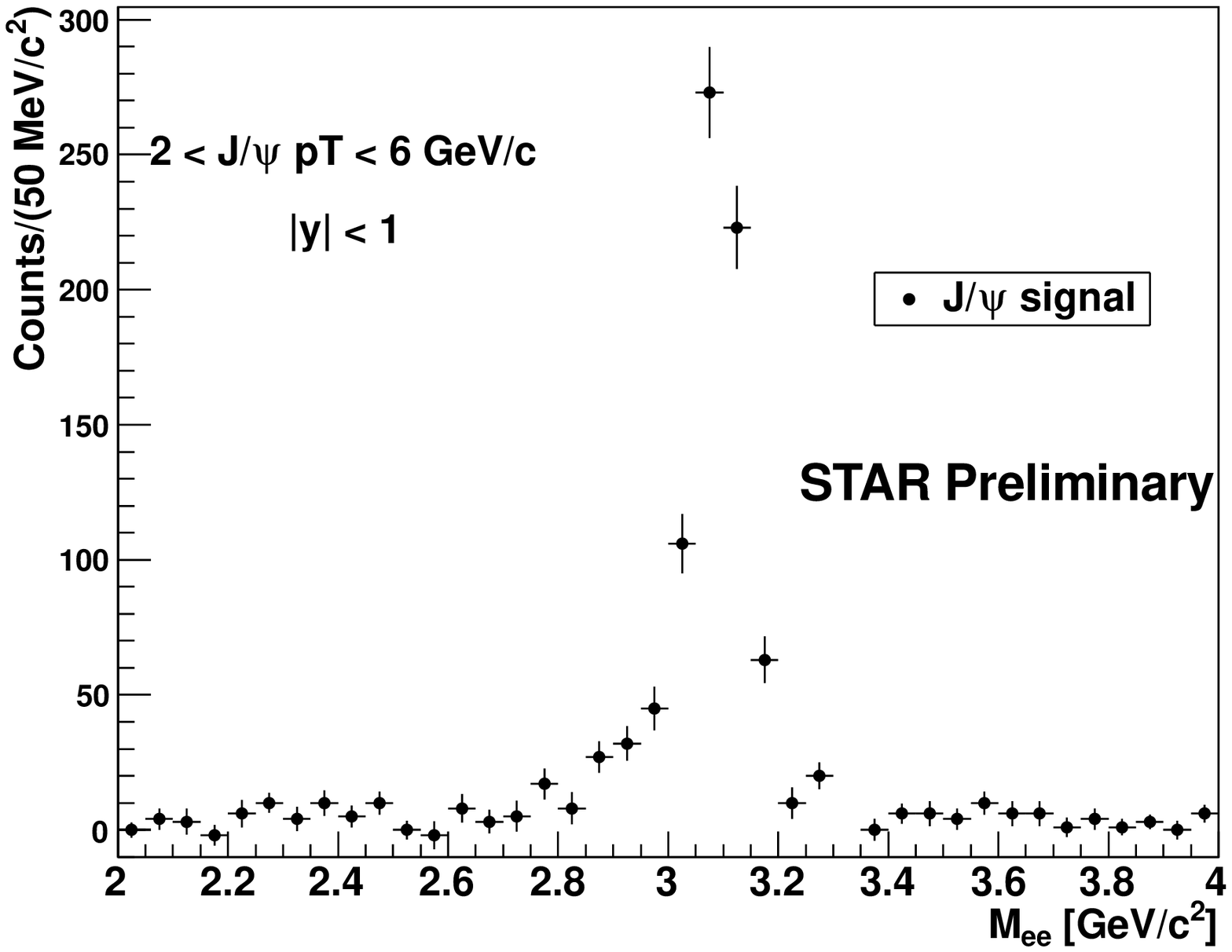} }
\label{massDistributions}
\caption[]{\footnotesize Invariant mass distributions in J/$\psi$ mass window (2.9 - 3.3) GeV/$c^2$, J/$\psi$ $p_T$ range (2 - 6) GeV/$c$ and $\mid y \mid < 1$. Fig. (a) shows signal+background in black (closed circles) and like-sign background in red (open circles). Fig. (b) shows  J/$\psi$ signal after the combinatorial background subtraction.}
 \vspace{-10pt}
\end{figure}

The cuts mentioned above allowed us to obtain a very clear  J/$\psi$ signal with a signal to background ratio of 15 and very high significance of 26$\sigma$. Fig. \ref{mass} shows the invariant mass distribution of all combinations of $e^+ e^-$ pairs in black (closed circles) and combinatorial background in red (open circles). The combinatorial background is calculated using like-sign technique, from a sum of $e^+ e^+$ and $e^- e^-$ pairs. The J/$\psi$ signal is obtained by subtracting the red histogram (background) from the black histogram (signal+background), see Fig. \ref{signal}. The number of  J/$\psi$'s calculated by counting bin entries in the invariant mass range (2.9 - 3.3) GeV/$c^2$ is 772 $\pm$ 29 in the J/$\psi$ $p_T$ range (2 - 6) GeV/$c$ and $\mid y \mid \;< $ 1. The same mass window is used for the polarization analysis for which the signal is split into 3 statistically comparable $p_T$ bins.

\section{J/$\psi$  polarization}
\begin{figure}[!ht]
 \vspace{-10pt}
\centering
\subfloat[\footnotesize J/$\psi$ $p_T$ range 2 $< p_T<$ 3 GeV/$c$]{\includegraphics[scale=0.25]{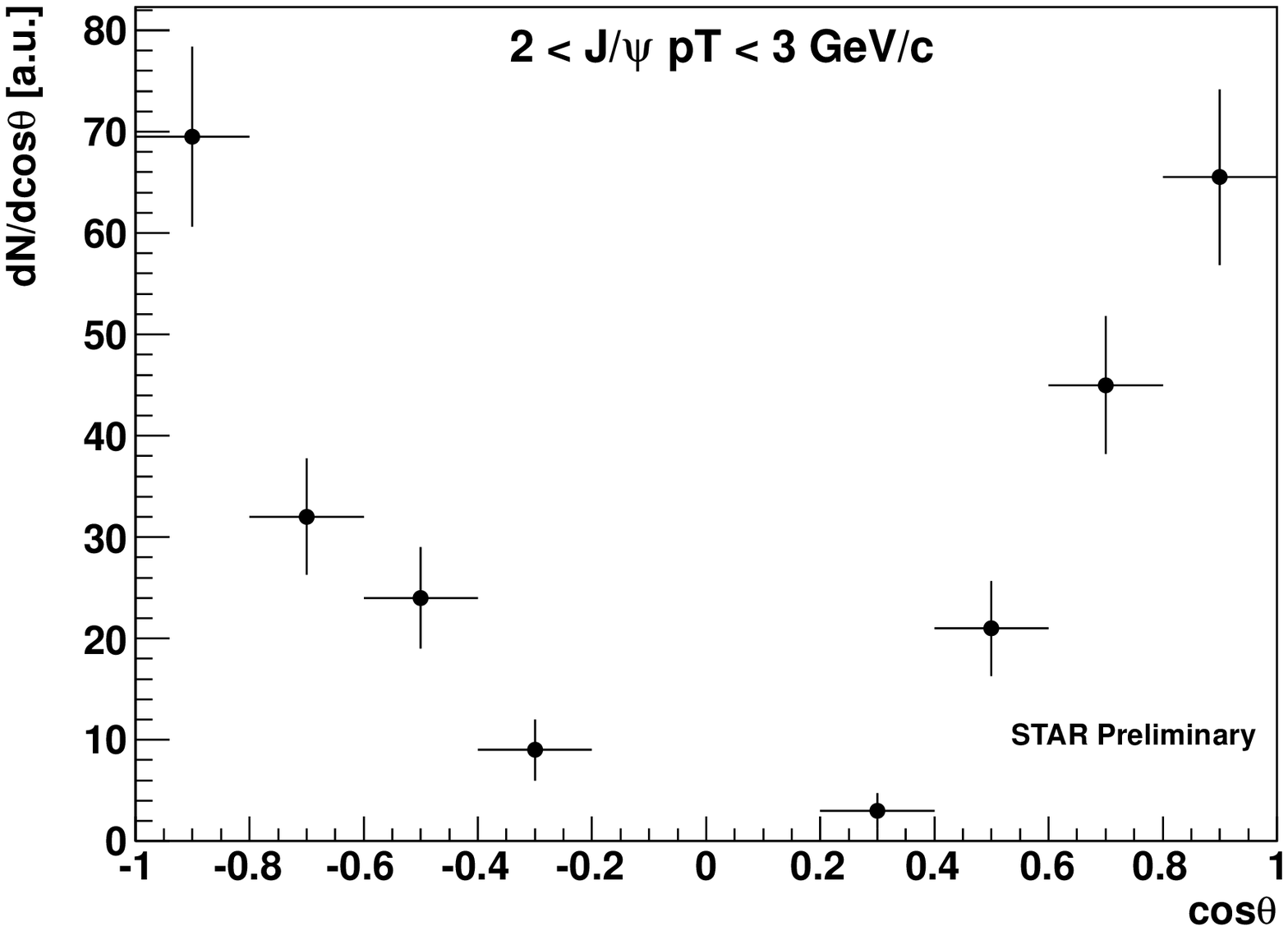} \label{rawCos1} }
\hspace{10 pt}
\subfloat[\footnotesize J/$\psi$ $p_T$ range 2 $< p_T<$ 3 GeV/$c$]{\includegraphics[scale=0.25]{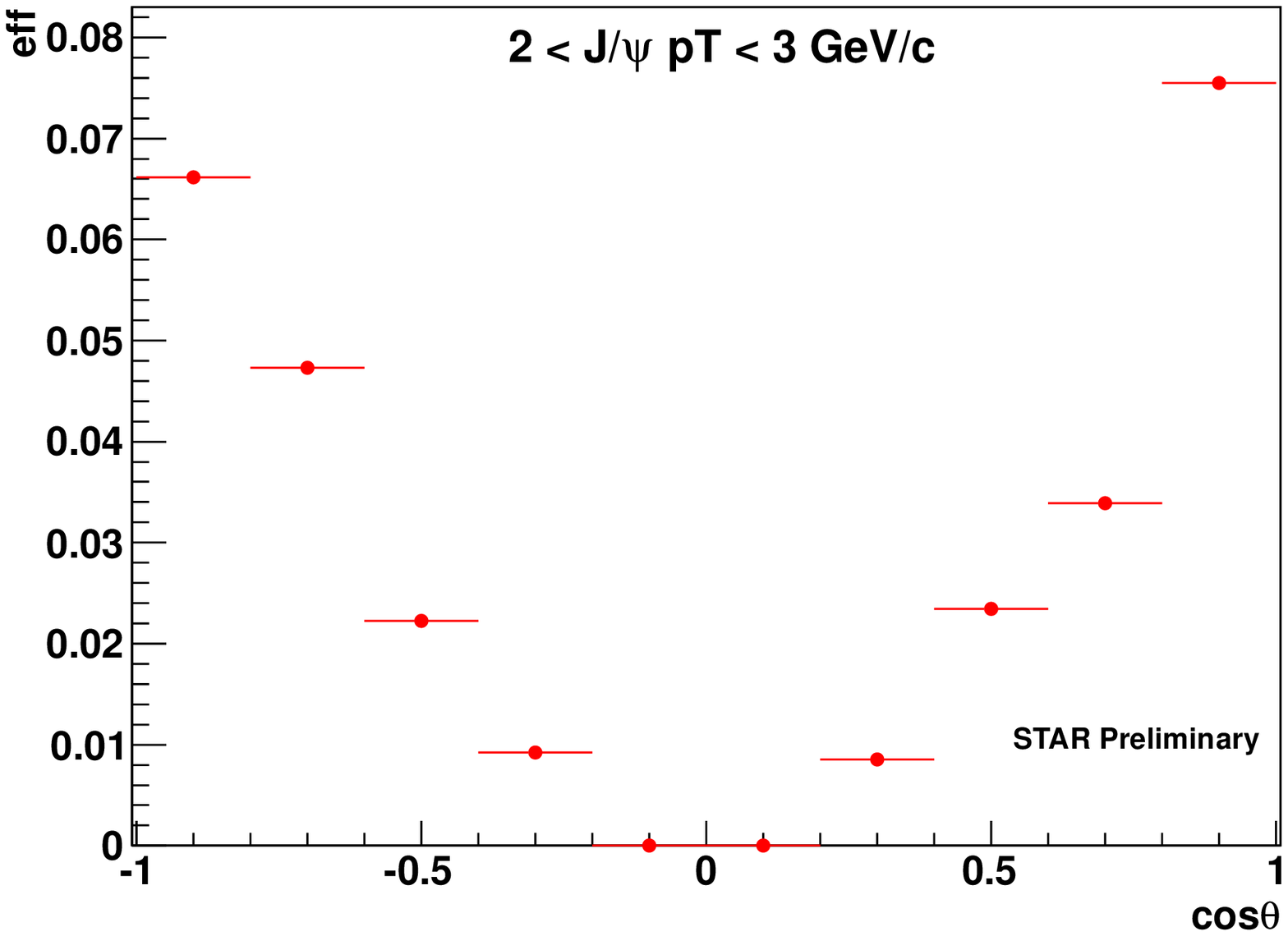} 
\label{corr1} }
\\  \vspace{-10pt}
\subfloat[\footnotesize  J/$\psi$ $p_T$ range 3 $< p_T<$ 4 GeV/$c$]{\includegraphics[scale=0.25]{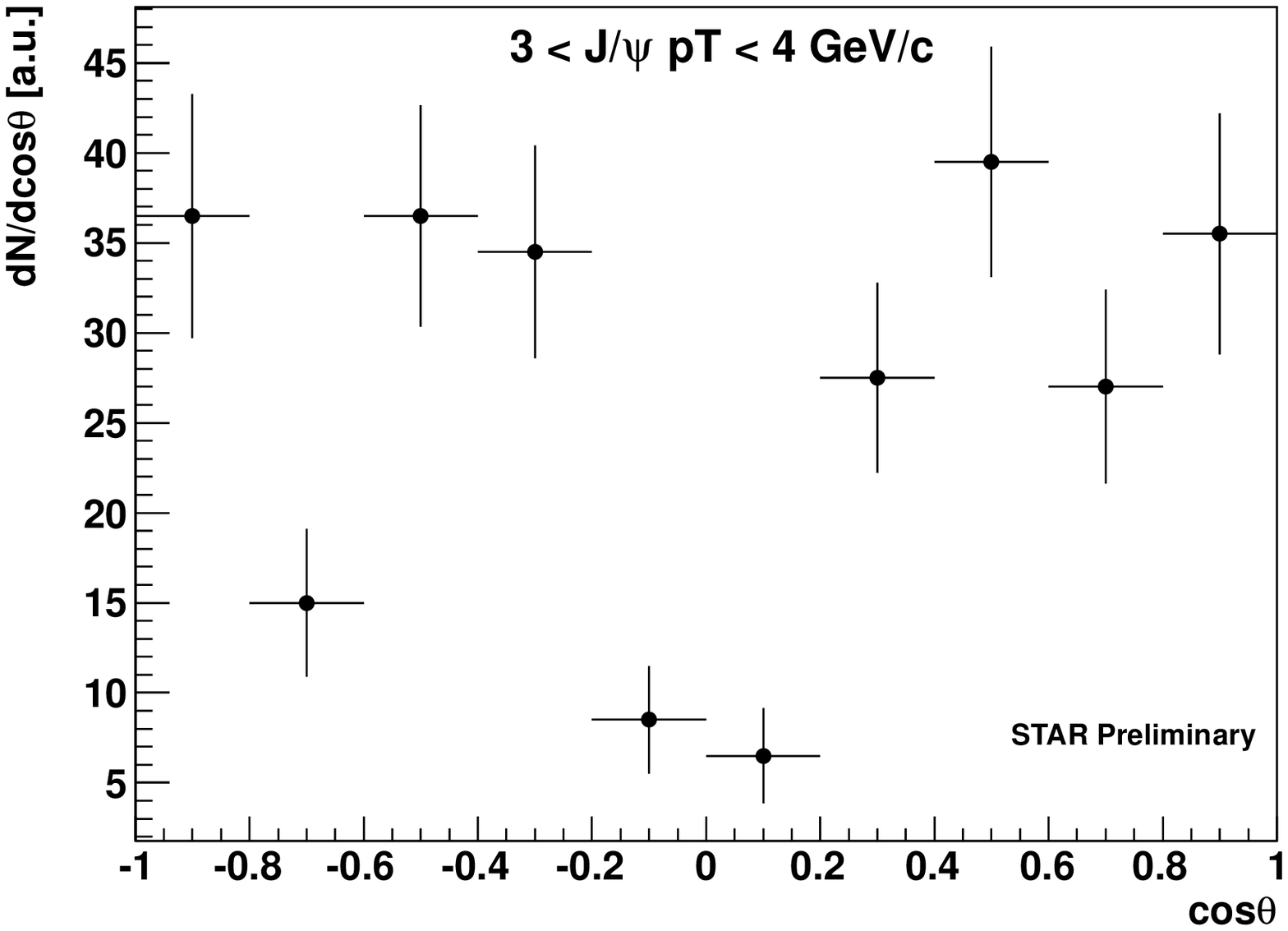} \label{rawCos2} }
\hspace{10 pt}
\subfloat[\footnotesize  J/$\psi$ $p_T$ range 3 $< p_T<$ 4 GeV/$c$]{\includegraphics[scale=0.25]{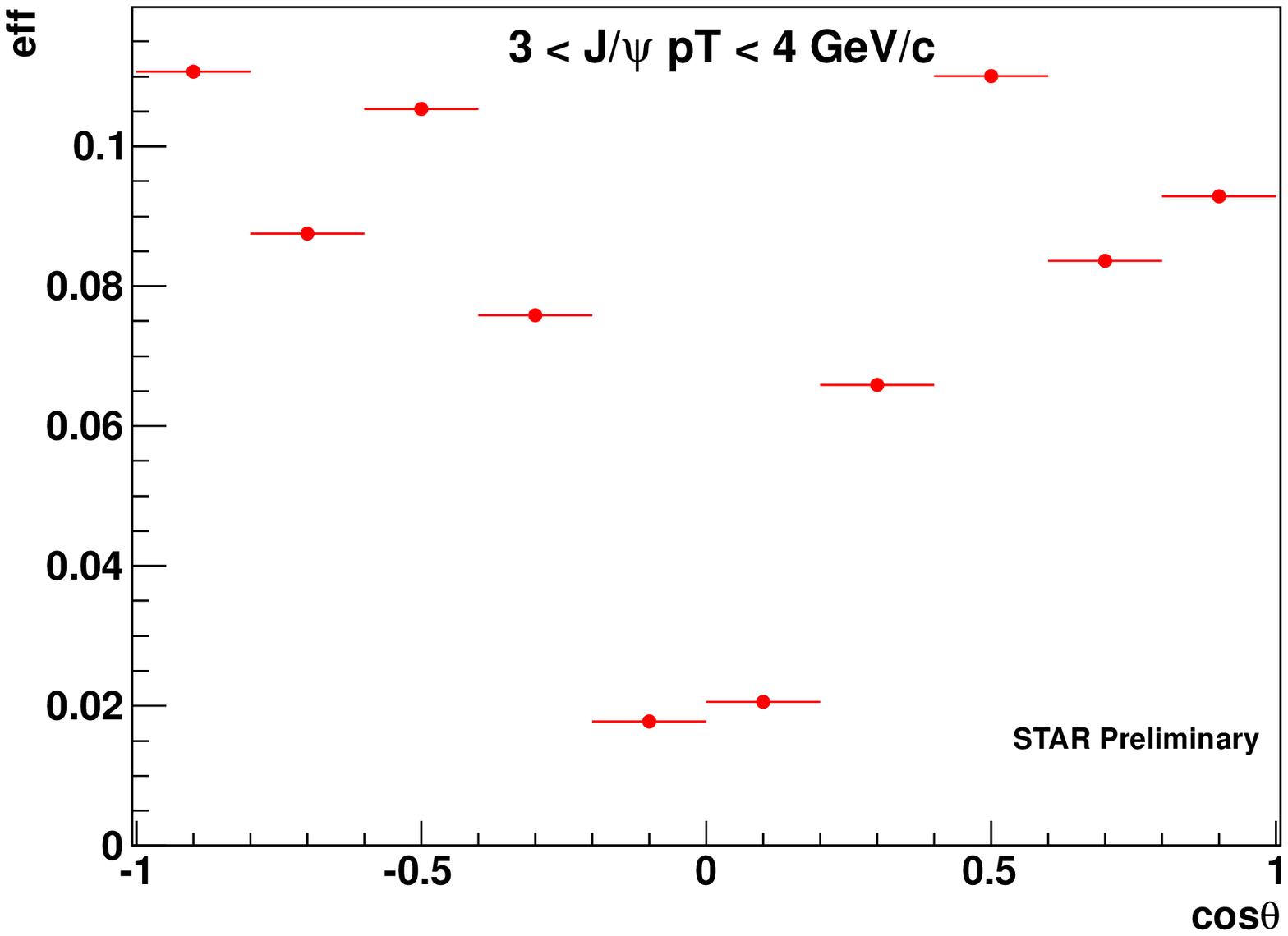} 
\label{corr2} }
\\ \vspace{-10pt}
\subfloat[\footnotesize  J/$\psi$ $p_T$ range 4 $< p_T<$ 6 GeV/$c$]{\includegraphics[scale=0.25]{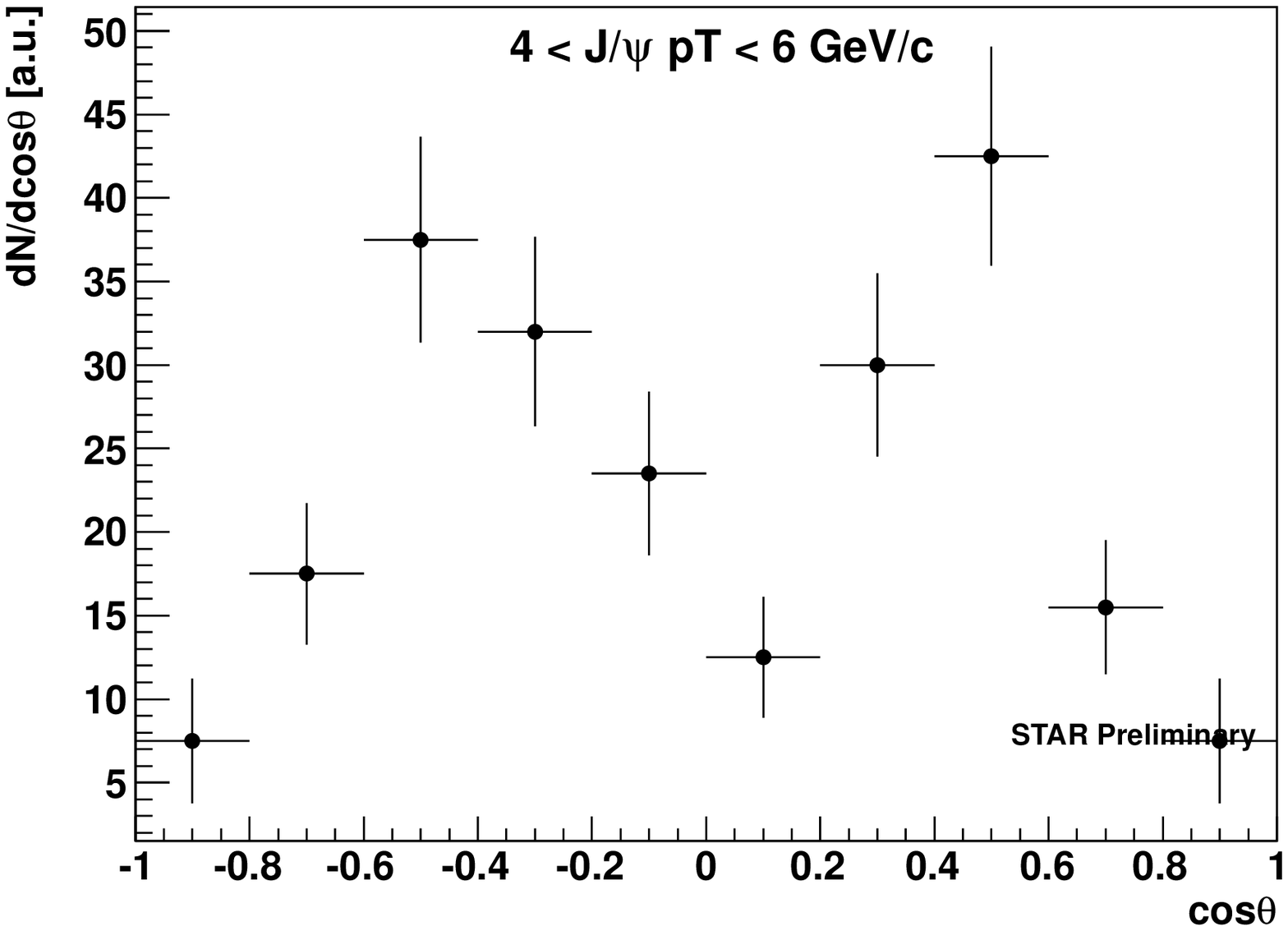} \label{rawCos3} }
\hspace{10 pt}
\subfloat[\footnotesize  J/$\psi$ $p_T$ range 4 $< p_T<$ 6 GeV/$c$]{\includegraphics[scale=0.25]{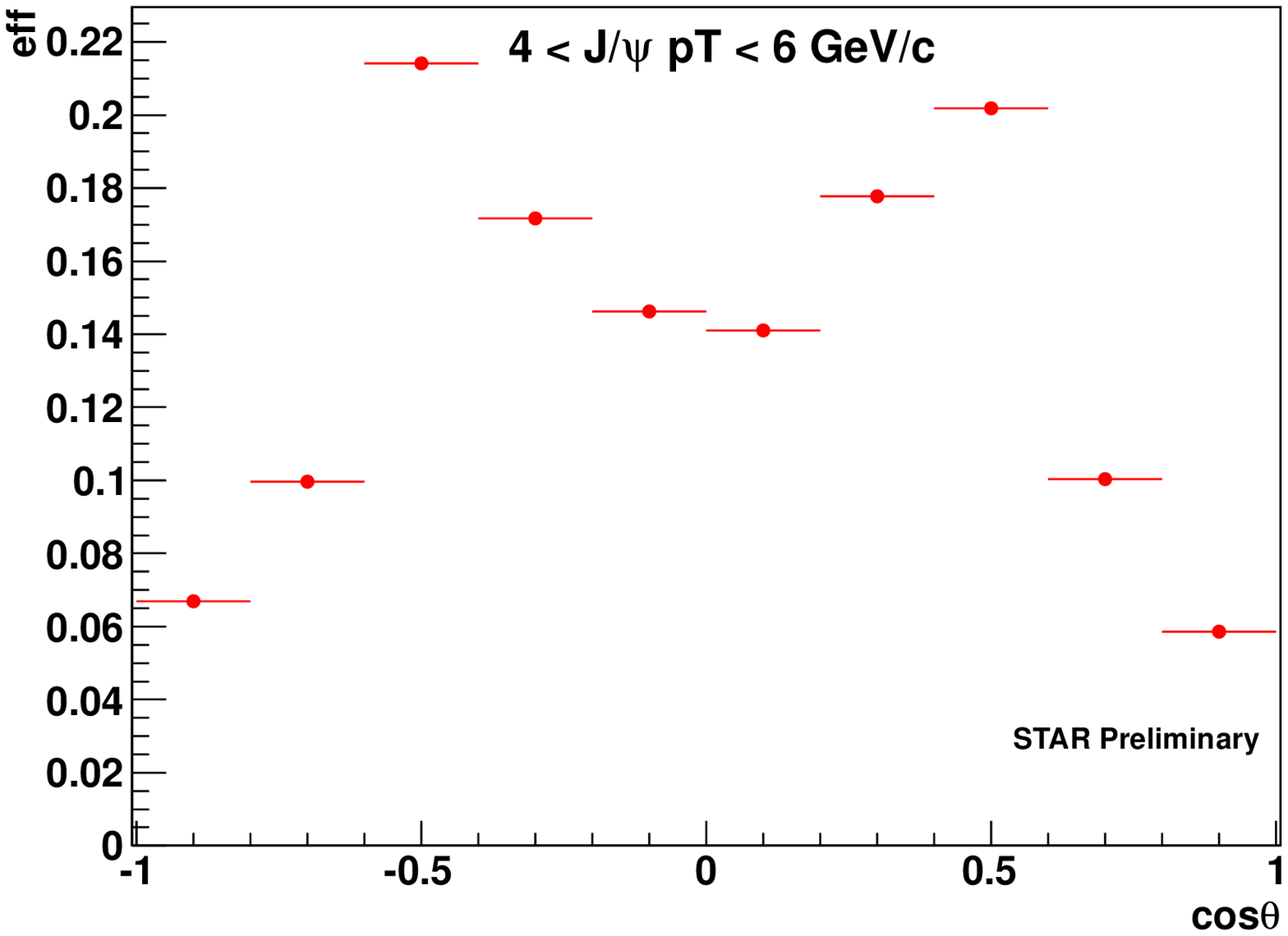} \label{corr3} }
\label{rawCos_corrections}
\caption[]{\footnotesize Left plots - uncorrected $cos \theta$ distribution with combinatorial background subtracted. Right plots - $cos \theta$ efficiency.}
 \vspace{-10pt}
\end{figure}

The uncorrected $cos\theta$ distribution is obtained using the same electron identification cuts as J/$\psi$ signal, in the J/$\psi$ $p_T$ range of (2 - 6) GeV/$c$ and  $\mid y \mid \;< $ 1. The $cos\theta$ distribution is divided into 3 J/$\psi$ $p_T$ bins: 2 $< p_T< 3$ GeV/$c$, 3 $< p_T<$ 4 GeV/$c$ and 4 $< p_T<$ 6 GeV/$c$. The $cos\theta$ distributions for the 3 $p_T$ bins, with combinatorial background subtracted are presented in Fig. \ref{rawCos1}, \ref{rawCos2}, \ref{rawCos3} respectively.

\subsection{Corrections}
In order to get the $cos\theta$ corrections, unpolarized Monte Carlo J/$\psi$'s with uniform $p_T$ and rapidity distributions are embedded into real events and the detector response is simulated. All cuts used in the analysis were applied and the $cos\theta$ efficiency as a function of J/$\psi$ transverse momentum was obtained. $cos\theta$  distributions were also re-weighted according to the real J/$\psi$ $p_T$ and y shapes. Obtained in that way, corrections were applied to uncorrected $cos\theta$ distributions from data in 1 GeV/$c$ wide J/$\psi$ $p_T$ bins. Corrections include acceptance correction, tracking efficiency, electron identification efficiency and trigger efficiency. The total J/$\psi$ efficiency is shown vs J/$\psi$ transverse momentum in Fig. \ref{corr1}, \ref{corr2}, \ref{corr3}. The most critical factor is trigger efficiency. At least one of electrons from J/$\psi$ decay must have $p_T$  above 2.6 GeV/$c$ since is required to satisfy the trigger conditions. It causes significant loss in number of observed J/$\psi$ at lower J/$\psi$ $p_T$ and the efficiency decrease with decreasing $\mid cos\theta \mid$. It is well visible in Fig. \ref{corr1}, where we lose all entries at $cos\theta \sim \;$ 0. With increasing J/$\psi$  $p_T$ the trigger efficiency increase but because the trigger has also the upper threshold ($E_{T} \; \leq$ 4.3 GeV) and due to the acceptance effect (single electron  $p_T > $ 0.4 GeV/$c$ and $\mid \eta \mid <$ 1) we see drop of total efficiency at $\mid cos\theta \mid \; \sim \;$ 1 , see Fig. \ref{corr3}.

\begin{figure}[!ht]
\centering
 \vspace{-10pt}
\subfloat[\footnotesize $\lambda$ = 0.46 $\pm$ 0.40 (stat.) $\pm$ 0.34 (sys.), 2 $< J/\psi \: p_T<$ 3 GeV/$c$]{ \includegraphics[scale=0.25]{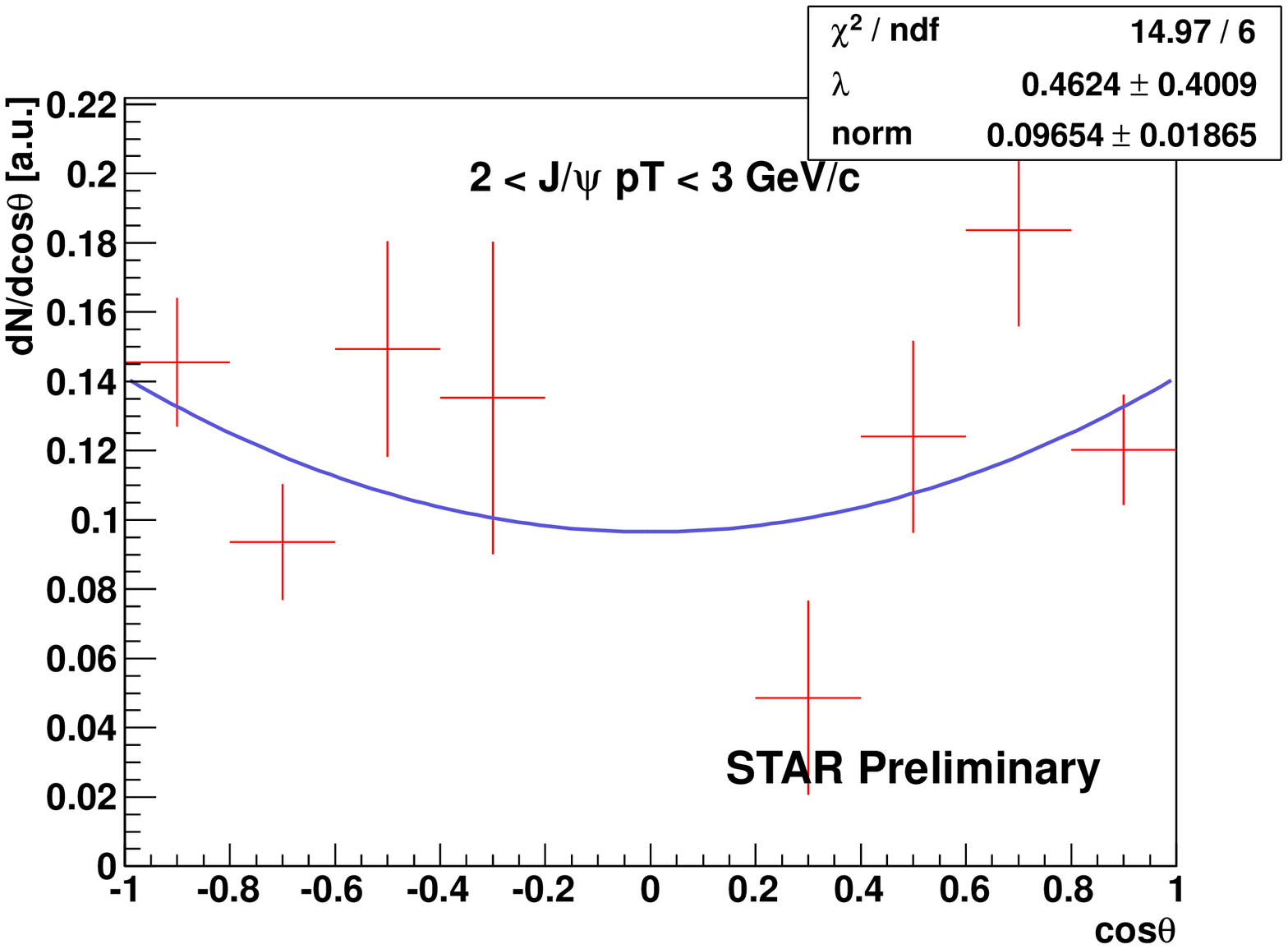} \label{cos1} }
\hspace{10 pt}
\subfloat[\footnotesize $\lambda$ = -0.32 $\pm$ 0.20 (stat.) $\pm$ 0.16 (sys.), 3 $< J/\psi \: p_T<$ 4 GeV/$c$]{ \includegraphics[scale=0.25]{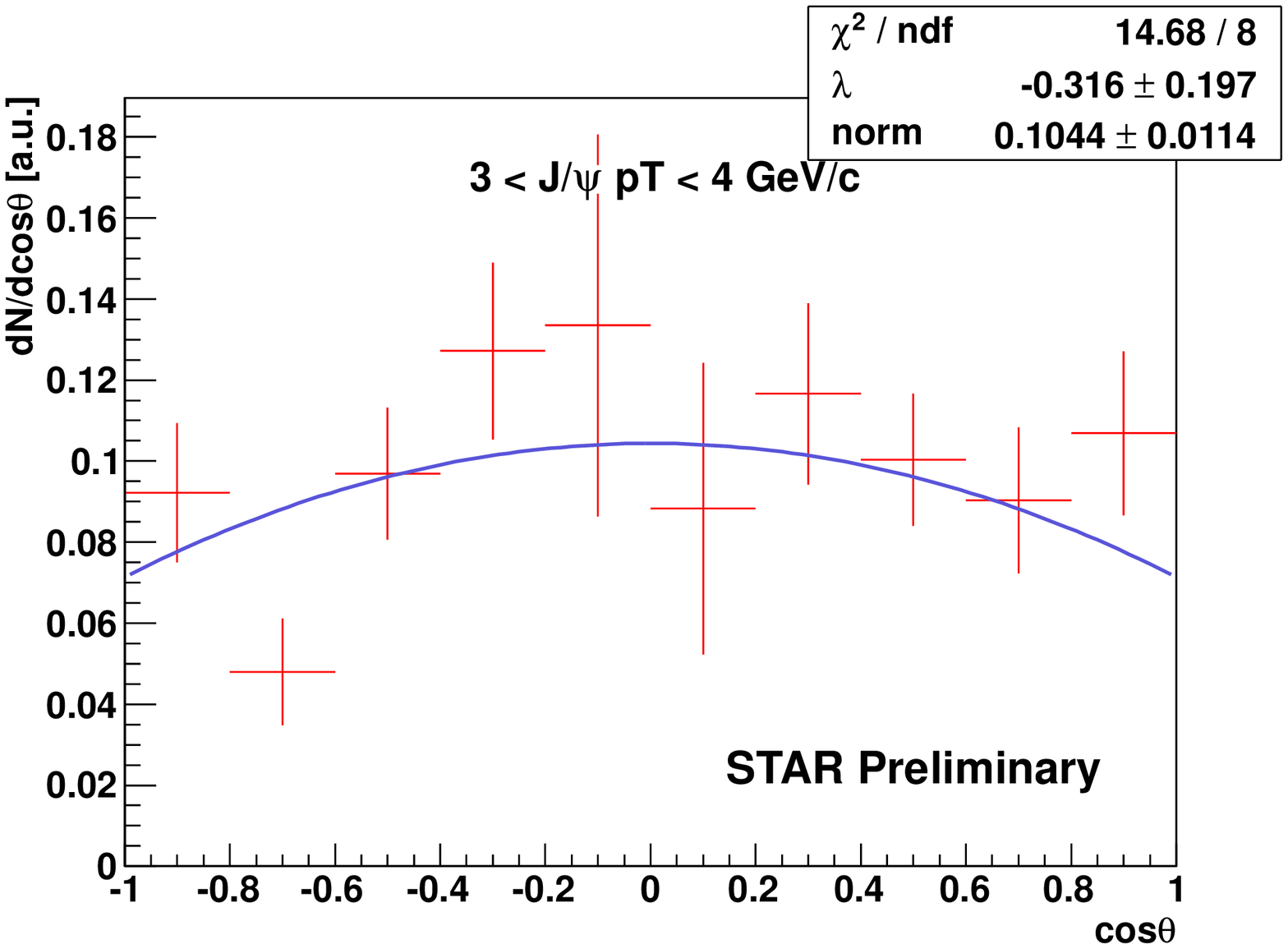} \label{cos2} }
\\
\subfloat[\footnotesize $\lambda$ = -0.17 $\pm$ 0.29 (stat.) $\pm$ 0.11 (sys.), 4 $< J/\psi \: p_T<$ 6 GeV/$c$]{ \includegraphics[scale=0.25]{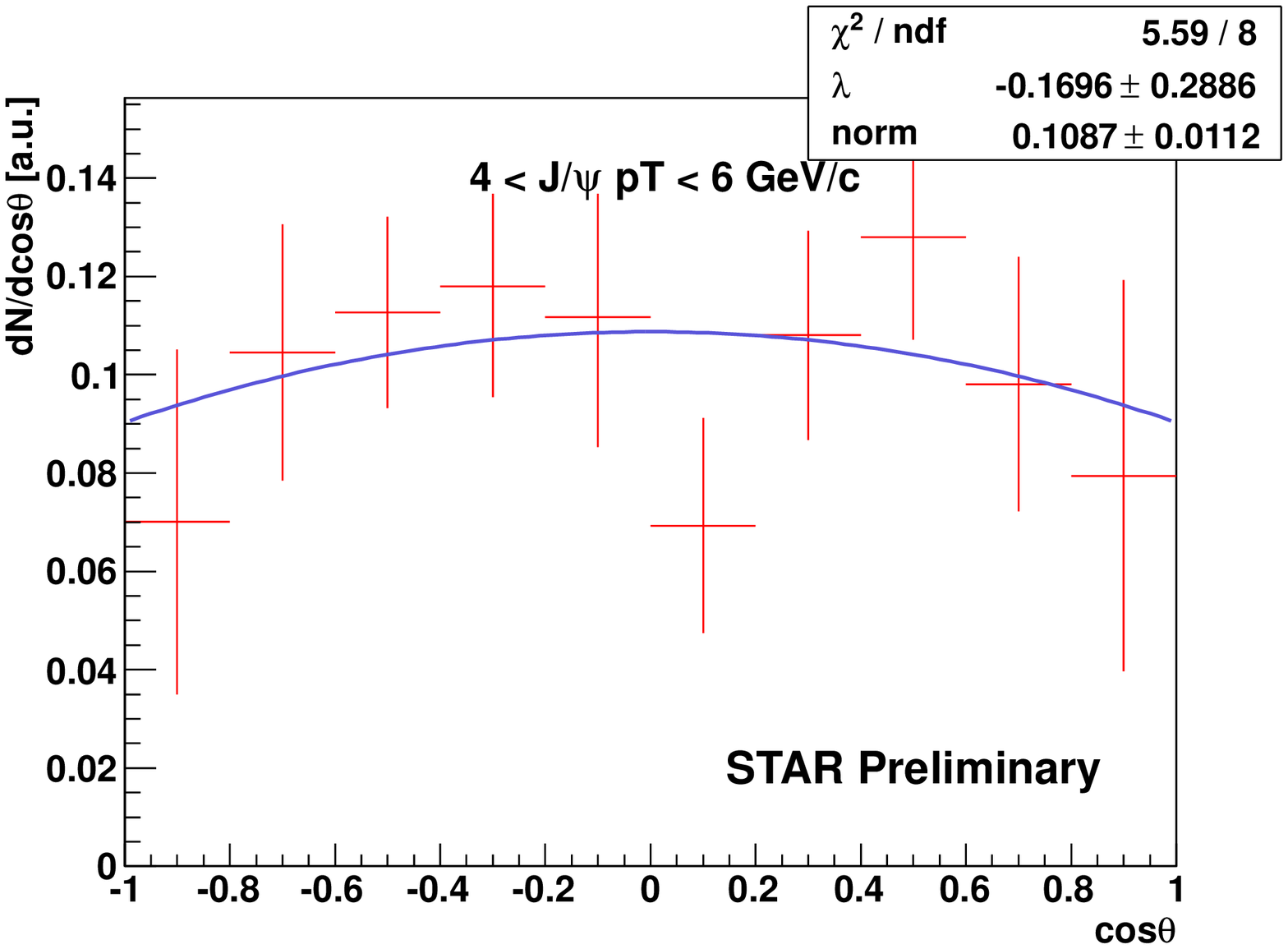} \label{cos3} }
\label{corrCos}
\caption[]{\footnotesize Corrected $cos \theta$ distributions with the fit: $norm (1+\lambda cos^2\theta)$,  errors are statistical}
 \vspace{-20pt}
\end{figure}

\subsection{ J/$\psi$ polarization result}
The corrected $cos \theta$ distributions are fitted with: $norm (1+\lambda cos^2\theta)$ (see Eq. \ref{parametrization}) with no constraints on the fit parameters, see Fig. \ref{cos1}, \ref{cos2}, \ref{cos3}, where $norm$ is a normalization factor, $\lambda$ is the polarization parameter. Lines represent the most likely fits. Obtained results of the polarization parameter are: $\lambda$ = 0.46 $\pm$ 0.40 (stat.) $\pm$ 0.34 (sys.), $\lambda$ = -0.32 $\pm$ 0.20 (stat.) $\pm$ 0.16 (sys.), $\lambda$ = -0.17 $\pm$ 0.29 (stat.) $\pm$ 0.11 (sys.) for  J/$\psi$ $p_T$ ranges: (2 - 3) GeV/$c$, (3 - 4) GeV/c, (4 - 6) GeV/$c$ respectively. 
Dominant sources of systematic uncertainties are: $cos \theta$ binning and acceptance, J/$\psi$ mass range, electron identification cut, trigger efficiency. Detailed work on systematic error estimation is in progress.

\begin{figure}[!ht]
\vspace{-20pt}
\centering
\includegraphics[width=0.75\textwidth]{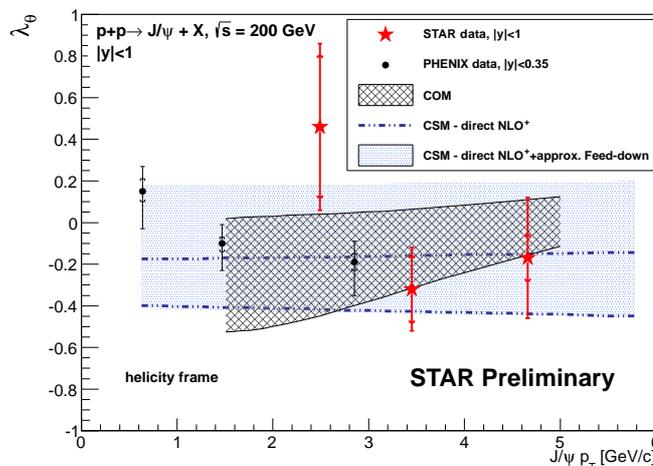}
 \vspace{-10pt}
\caption{\footnotesize STAR polarization parameter $\lambda$ vs J/$\psi$ $p_T$ (red star symbols) compared with PHENIX result (black symbols) \cite{PHENIX}, $NLO^+ CSM$ (blue shaded area) \cite{Lansberg} and $COM$ \cite{NRQCD} (gray hatched area) predictions. }
\label{polarization}
 \vspace{-10pt}
\end{figure}

Polarization parameters as a function of  J/$\psi$ $p_T$ are shown in Fig. \ref{polarization}. The STAR result (red star symbols) is compared with $NLO^+ CSM$ \cite{Lansberg} (blue shaded area) and $COM$ \cite{NRQCD} (gray hatched area) model predictions and with the PHENIX result for J/$\psi$ polarization at mid-rapidity (black symbols) \cite{PHENIX}. The STAR result is consistent with the COM and CSM predictions within current experimental and theoretical uncertainties. The measurement is also consistent with the PHENIX data and extends the $p_T$ reach to $\sim$ 6 GeV/$c$.

\section{ Summary}
In this report, the J/$\psi$ polarization measurement from the STAR experiment at mid-rapidity is presented. The polarization parameter $\lambda$ is extracted in the helicity frame in 3 J/$\psi$ $p_T$ bins. Within current uncertainties the obtained transverse momentum dependent $\lambda$ parameter is consistent with $NLO^+ CSM$ and $COM$ model predictions, and with no polarization.

\begin
{thebibliography}{999}
\bibitem{Lansberg}J.P. Lansberg, Phys. Lett. B 695 (2011) 149-156
\bibitem{CDF}A.Abulencia et al. (CDF Collaboration), Phys. Rev. Lett. 99, 132001 (2007)
\bibitem{NRQCD}H.S. Chung, C. Yu, S. Kim, J. Lee, Phys. Rev. D 81, 014020 (2010)
\bibitem{HX}C.S. Lam, W.K. Tung, Phys. Rev. D 18, 2447 (1978)
\bibitem{PHENIX}A. Adare et al. (PHENIX Collaboration), Phys. Rev. D 82, 012001 (2010) 
\end{thebibliography}

\end{document}